\documentclass[%
 aip, amsmath,amssymb,
floatfix,%
reprint,%
]{revtex4-1}

\usepackage{graphicx}
\usepackage{dcolumn}
\usepackage{bm}

\usepackage[utf8]{inputenc}
\usepackage[T1]{fontenc}
\usepackage{mathptmx}
\usepackage{etoolbox}
\usepackage{multirow}
\usepackage{hyperref}

\makeatletter
\def\@email#1#2{%
 \endgroup
 \patchcmd{\titleblock@produce}
  {\frontmatter@RRAPformat}
  {\frontmatter@RRAPformat{\produce@RRAP{*#1\href{mailto:#2}{#2}}}\frontmatter@RRAPformat}
  {}{}
}%
\makeatother
\begin{document}

\title{Delocalized Electronic Excitations and their Role in Directional Charge Transfer in the Reaction Center of Rhodobacter Sphaeroides}

\author{Sabrina Volpert}
\affiliation{Institute of Physics, University of Bayreuth, 95440 Bayreuth, Germany}
\author{Zohreh Hashemi}
\affiliation{Institute of Physics, University of Bayreuth, 95440 Bayreuth, Germany}
\author{Johannes M. Foerster}
\affiliation{Institute of Physics, University of Bayreuth, 95440 Bayreuth, Germany}
\author{Mario R. G. Marques}
\affiliation{Institute of Physics, University of Bayreuth, 95440 Bayreuth, Germany}
\author{Ingo Schelter}
\affiliation{Institute of Physics, University of Bayreuth, 95440 Bayreuth, Germany}
\author{Stephan K\"ummel}
\affiliation{Institute of Physics, University of Bayreuth, 95440 Bayreuth, Germany}
\author{Linn Leppert}
\affiliation{MESA+ Institute for Nanotechnology, University of Twente, 7500 AE Enschede, The Netherlands}
\affiliation{Institute of Physics, University of Bayreuth, 95440 Bayreuth, Germany}
\email{l.leppert@utwente.nl}

\date{\today}

\begin{abstract}
In purple bacteria, the fundamental charge-separation step that drives the conversion of radiation energy into chemical energy proceeds along one branch - the A branch - of a heterodimeric pigment-protein complex, the reaction center. Here, we use first principles time-dependent density functional theory (TDDFT) with an optimally-tuned range-separated hybrid functional to investigate the electronic and excited-state structure of the primary six pigments in the reaction center of \textit{Rhodobacter sphaeroides}. By explicitly including amino-acid residues surrounding these six pigments in our TDDFT calculations, we systematically study the effect of the protein environment on energy and charge-transfer excitations. Our calculations show that a forward charge transfer into the A branch is significantly lower in energy than the first charge transfer into the B branch, in agreement with the unidirectional charge transfer observed experimentally. We further show that inclusion of the protein environment redshifts this excitation significantly, allowing for energy transfer from the coupled $Q_x$ excitations. Through analysis of transition and difference densities, we demonstrate that most of the $Q$-band excitations are strongly delocalized over several pigments and that both their spatial delocalization and charge-transfer character determine how strongly affected they are by thermally-activated molecular vibrations. Our results suggest a mechanism for charge-transfer in this bacterial reaction center and pave the way for further first-principles investigations of the interplay between delocalized excited states, vibronic coupling, and the role of the protein environment of this and other complex light-harvesting systems. 
\end{abstract}

\maketitle

\newpage

\section{Introduction}
\begin{figure}[htb]
	\centering
        \includegraphics[width=0.45\textwidth]{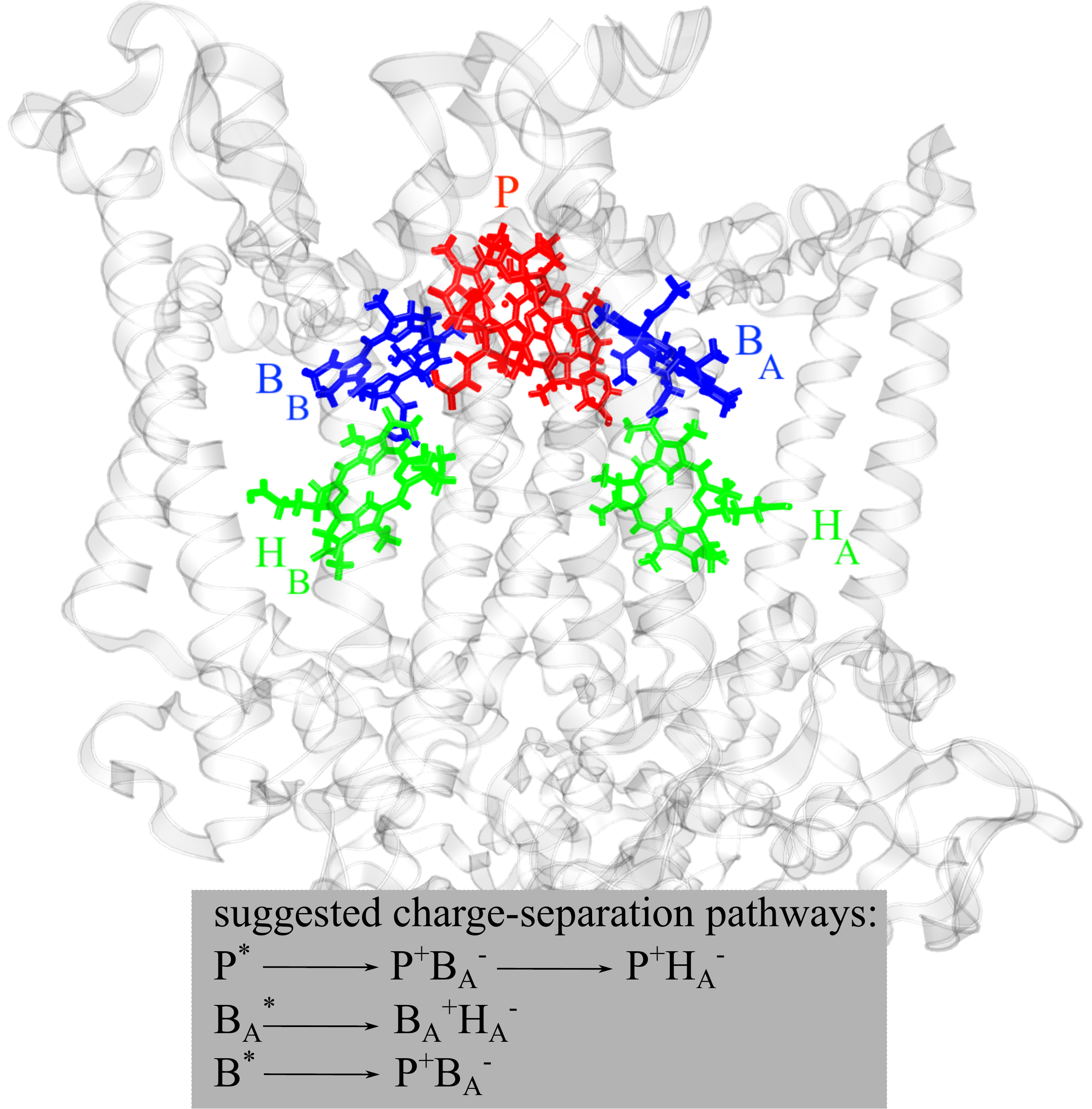}
		\caption{Proposed charge-separation pathways \cite{Niedringhaus2018, Ma2019, van_brederode_efficiency_1998, zhou_probing_1998, lin_excitation_1998, huang_cofactor-specific_2012} and structural model of the RC of \textit{Rhodobacter sphaeroides} including the special pair ($P$ in red), two accessory BCLs ($B_A$, $B_B$ in blue), and two bacteriopheophytines ($H_A$, $H_B$ in green). Protein chains are shown in transparent grey. Hydrogen atoms are omitted for clarity.}
\label{fig:structure}
\end{figure}

In natural photosynthesis the energy of sun light is converted into chemical energy in highly efficient excitation- and charge-transfer processes \cite{Renger2008}. Absorption of light happens primarily in antenna complexes, which funnel the excitation energy towards the reaction center (RC) where a charge-separation step initiates a cascade of electron-transfer processes resulting in a proton gradient that drives the biochemical reactions of photosynthesis. In purple bacteria such as \textit{Rhodobacter sphaeroides}, the fundamental design principles of these pigment-protein complexes are well understood due to a wealth of experimental and computational techniques that give access to detailed structural and spectroscopic information \cite{Jordanides2001,camara-artigas_interactions_2002,Jonas1996,Schlau-Cohen2012,Rancova2016, Niedringhaus2018,kawashima_energetic_2018,Kavanagh2020}. In this respect, the bacterical RC can also be understood as a model system for the RC of more complex photosynthetic organisms because its structure is highly conserved across bacteria, algae, and plantsc\cite{Cogdell2006}. Its main building blocks, shown in Figure~\ref{fig:structure}, are arranged along two pseudosymmetric branches $A$ and $B$ and consist of a strongly coupled dimer of two bacteriochlorophyll (BCL) molecules dubbed the special pair ($P$), two accessory BCLs ($B_A$, $B_B$), two bacteriopheophytines ($H_A$, $H_B$), and two quinones ($Q_A$, $Q_B$) embedded in a transmembrane protein matrix.

Despite the similar, but not exactly symmetric, structure of the $A$ and $B$ branches, it is well-established that the primary charge separation reaction only proceeds along the $A$ branch with near-unity quantum efficiency \cite{wraight_absolute_1974, kirmaier_picosecond-photodichroism_1985, zinth_first_2005}. The ultrafast timescales on which the primary energy- and charge-transfer processes occur in the RC in combination with broad overlapping absorption peaks originating from the coupling of multiple pigments and their protein environments, have posed significant challenges to spectroscopic techniques. Two-dimensional electronic spectroscopy (2DES) has become one of the primary experimental techniques for studying the bacterial RC \cite{ma_vibronic_2018,Niedringhaus2018,Ma2019,policht_hidden_2022} and other photosynthetic systems \cite{tros_complete_2020, nguyen_charge_2022}. For the RC of \textit{Rhodobacter sphaeroides} and similar RCs, 2DES has been used to propose models for the kinetics of the primary charge separation. In these models, it is usually assumed that charge separation is initiated through the excitation of $P$ (denoted as $P^*$), leading to a charge-transfer intermediate $P_A^+P_B^-$, which is followed by a charge-separated state $P^+H_A^-$ via an ultra-shortlived intermediate $P^+B_A^-$ \cite{Niedringhaus2018, Ma2019}, however, alternative charge-separation pathways have been suggested as well, for example starting from an excitation localized on one or both accessory BCLs, i.e., $B_A^*$ or $B^*$ \cite{van_brederode_efficiency_1998, zhou_probing_1998, lin_excitation_1998, huang_cofactor-specific_2012} (Figure~\ref{fig:structure}).

Computational modelling has played an important role in helping to unravel the intricate factors affecting excitation and charge transfer in the antenna complexes and the RC. Calculations based on model Hamiltonians and advances in the semiempirical modelling of long- and short-range coupling between chromophores have allowed for the simulation of excitation processes in large pigment-protein complexes \cite{Madjet2009a,Zheng2016,Jang2018}. However, also computationally more demanding first-principles calculations, primarily based on (time-dependent) density functional theory (TDDFT), have been employed to study systems of growing complexity and size. The focus of many studies has been the origin of the unidirectionality of the charge-separation process in bacterial RCs and photosystem II of plants. 

For the RC of photosystem II, a large model system was used by T.~Frankcombe including four (truncated) chlorophyll, two pheophytin, and two plastoquinone molecules using TDDFT and a polarizable continuum model (PCM) to account for effects of the protein environment \cite{Frankcombe2015a}. Later, Sirohiwal \textit{et al.} reported TDDFT calculations using a range-separated hybrid functional on chlorophyll monomers, dimers, and trimers of the photosystem II RC showing that the lowest-energy charge-transfer excitation corresponds to $B_A^+H_A^-$ and is strongly affected by protein electrostatics \cite{Sirohiwal2020c}. Low-energy charge-transfer excitations were also reported by Kavanagh \textit{et al.} in TDDFT calculations using a hexameric model of the photosystem II RC including parts of the protein environment explicitly. Similarly, Förster \textit{et al.} observed an excitation with partial $H_A^-B_A^+$ charge-transfer character using the $GW$+Bethe-Salpeter Equation approach \cite{forster_quasiparticle_2022}. 

For the RC of \textit{Rhodobacter sphaeroides}, (TD)DFT calculations including the special pair and some of its neighboring amino-acid residues were reported in 2011 by Wawrzyniak \textit{et al.}~and indicated that protein induced distortions of the special pair geometry lead to an asymmetric ground-state electron density \cite{Wawrzyniak2011}. A similar model system was employed by Eisenmayer \textit{et al.}~, who performed molecular dynamics simulations based on constrained DFT and showed that the electron-density asymmetry is dynamical and coupled to a low-frequency vibrational mode, related to the rotation of a histidine residue close to $B_A$ \cite{Eisenmayer2012}. Later, Eisenmayer \textit{et al.}~included the $B_A$ in their constrained DFT simulations showing the coupling of proton displacements to the primary electron-transfer step from $P$ to $B_A$ \cite{Eisenmayer2013}. Aksu \textit{et al.}~combined TDDFT with a tuned range-separated hybrid functional with PCM and showed that spectral asymmetries arise from locally different dielectric environments along the $A$ and $B$ branches \cite{Aksu2019}. The initial charge-transfer excitations of $P$ were also studied by Aksu \textit{et al.}~employing the same methodology \cite{Aksu2020}. (TD)DFT calculations by Mitsuhashi \textit{et al.} in which the environment of $P$ together with either $B_A$ or $B_B$ was represented using a QM/MM/PCM scheme, further indicated that the lowest unoccupied molecular orbital (LUMO) of $B_A$ is lower in energy than the LUMO of $B_B$, suggesting that $B_A$ is the primary electron acceptor \cite{Mitsuhashi2021}. Another study in which some of the same authors used a diabatization scheme to evaluate electronic couplings between $P$ and $B_A$ and $B_B$, respectively, pointed to the particular importance of a tyrosene residue close to $B_A$ as being responsible for the directionality of charge transfer \cite{Tamura2021}. Br\"utting \textit {et al.} investigated the primary charge separation step in the quasi-symmetric reaction center of \textit{Heliobacterium Modesticaldum}, with an emphasis on revealing the influence of nuclear motion on the relative energetic positions of different electronic excitations \cite{Brutting2021}.

To the best of our knowledge, explicit TDDFT calculations on a reaction center model of \textit{Rhodobacter sphaeroides} including all six primary pigments and parts of the environment have not been reported yet. Furthermore, while previous studies have provided detailed insight into the effects of the protein environment and molecular vibrations on excited states, little attention has been directed at the delocalized, correlated multi-particle nature of these excitations. One may wonder whether these characteristics can be properly captured by TDDFT, as the wave-functions obtained in TDDFT have no rigorous physical meaning. We here show, however, that one can analyze the excitations reliably based on transition densities and difference densities, i.e., quantities that have a solid foundation in TDDFT. 

To this end, we use TDDFT with an optimally-tuned range-separated hybrid functional to study a hexameric model of the RC including the primary pigments, i.e., the special pair $P$, the accessory BCLs $B_A$ and $B_B$ and the bacteriopheophytins $H_A$ and $H_B$. We also explicitly model the effect of close-lying amino-acid residues on the excited states by including them in our TDDFT calculations. We clarify which amino acids are responsible for significant changes in excited-state energies and compare our results with QM/MM calculations. Our calculations show that a distinction between localized excitations on the one hand and charge-transfer excitations on the other hand is of limited usefulness to understand the excited state structure of this system of strongly coupled pigments. Instead, we find excitations without charge-transfer character that are delocalized across several pigments and that can not readily be classified as coupled excitations of individual monomeric units. Partial charge-transfer states between the special pair pigments ($P_A^+P_B^-$) are low in energy, mix with these delocalized states and are a consequence of the strong coupling between the pigments. The lowest-energy charge-transfer state that transfers an electron into the A branch can clearly be classified as $B_A^+H_A^-$. This is in agreement with previous first-principles calculations on the photosystem II reaction center, but not in line with experimental reports suggesting charge-transfer through an intermediate $P^+B_A^-$ states. The $B_A^+H_A^-$ excitation is $\sim$20\,meV higher in energy than the highest-energy $Q$-band excitations. Although we cannot rule out that including further parts of the environment might lower its energy further, such a small energetic separation suggests that the vibrational modes of the pigments and/or the environment could couple this charge-transfer state to the delocalized $Q$-band excitations.

\section{Computational Methods} \label{sec:methods}
\subsection{Structure of Model Systems}\label{sec:structures}
All our calculations are based on the experimental crystal structure of the wild-type RC of \textit{Rhodobacter sphaeroides} with Protein Data Bank file ID 1M3X \cite{camara-artigas_interactions_2002}. The pigment-protein complex has two main protein chains called L- and M-chains which form the backbone of the $A$ and $B$ branches, respectively. We are interested in the primary charge-transfer process and, therefore, included $P$, $B_A$, $B_B$, $H_A$, and $H_B$ in all our computational models. For approximating the effect of the protein environment on energy- and charge-transfer excitations, we added amino-acid residues explicitly to our model structures, described in more detail in Section~\ref{sec:tetrameric}. Hydrogen atoms are not resolved in the experimental crystal structure and are therefore added with the module \textsc{hbuild} in \textsc{charmm} \cite{Brooks2009} and energetically optimized using the \textsc{charmm} force field \cite{CHARMMFF} as described in Ref.~\citenum{Schelter2019}. In all model systems, we cropped the phytyl tails of the BCL molecules and saturated the carboxyl group with a hydrogen atom. Using a methyl group to saturate the phytyl tail does not change the main conclusions of this paper, as shown in Figure S1 of the Supplementary Material (SM). Furthermore, we cut the bonds between the amino-acid residues and the polypeptide chains between $C_\alpha$ and $C_\beta$ and saturated them with hydrogen atoms.

\subsection{TDDFT Calculations}\label{sec:TDDFT}
We performed (TD)DFT calculations using \textsc{q-chem}, version 5.2.2 \cite{Shao2015}, \textsc{turbomole} version 7.5 \cite{noauthor_turbomole_nodate, balasubramani_turbomole_2020}, and \textsc{orca} version 5.0.2 \cite{Neese2012}. We used the Pople basis set 6-31G(d,p) for which the $Q_y$ and $Q_x$ excitation energies of a single BCL~\textit{a} molecule are converged to within 50\,meV \cite{Schelter2019}. We also tested the accuracy of the basis set for the special pair $P$ as discussed in the SM (Table S1). The exchange-correlation energy is approximated using the optimally-tuned $\omega$PBE functional \cite{Vydrov2006} which has been shown to properly capture the coupling between BCLs \cite{Schelter2019} and to be \textit{on par} with Green's function-based many-body perturbation theory for a wide range of single chromophores \cite{DeQueiroz2021, Hashemi2021a}. Range-separated hybrid functionals have also been demonstrated to accurately describe electrochromic shifts due to the protein environment of various biochromophores in an extensive benchmark of DFT approximations by Sirohiwal \textit{et al.} \cite{Sirohiwal2021}. In the optimally-tuned $\omega$PBE functional, the range-separation parameter determines the length scale at which short-range semilocal exchange goes over into exact long-range exchange. Such functionals significantly improve the description of charge-transfer excitations \cite{Kuemmel2017} and lead to excellent agreement with experimental photoemission spectroscopy for a broad range of systems from molecules to solids \cite{Refaely-Abramson2011, Refaely-Abramson2012, Korzdorfer2012, Refaely-Abramson2013, DeQueiroz2014, manna_quantitative_2018, Wing2020a}. In the optimal-tuning procedure, the range-separation parameter $\omega$ is varied such that the difference between the HOMO eigenvalue $\varepsilon_{HOMO}$ and the negative ionization potential of both the neutral and the anionic system is minimized \cite{Kronik2012}. Here, we use $\omega=0.171a_0^{-1}$ based on tuning for one BCL~$a$ performed by Schelter \textit{et al.}\cite{Schelter2019}. We confirmed that the deviation of the ionization potentials from $-\varepsilon_{HOMO}$ of $P$ and of a single BCL~$a$ with coordinating histidine are negligible and do not perform a separate tuning procedure for each of our model systems. This approach is also supported by more general arguments: Using the same $\omega$ for each model system allows us to compare the electronic and excited state structure of our model systems on the same footing. Furthermore, optimal tuning of conjugated systems of increasing size leads to artificially low values of $\omega$ and, thus, a dominance of semilocal exchange at long range, which deteriorates the description of charge-transfer excitations \cite{Korzdorfer2011a, DeQueiroz2014}.
For our TDDFT calculations, we used the linear response Casida approach and did not make the Tamm-Dancoff approximation (TDA) unless otherwise noted. We provide further information regarding the numerical convergence of our calculations in the SM. Details of our QM/MM TDDFT calculations with \textsc{orca} can also be found in the SM.

\subsection{Classification of Charge-Transfer Excitations}\label{sec:CT}
Since the transition density vanishes for charge-transfer states, we calculated the difference density $\Delta n_i = n_i - n_0$ between the excited ($n_i$) and the ground-state density ($n_0$) for every excitation $i$. The excited-state density $n_i$ is calculated as the diagonal part of the excited state density matrix $\gamma^{ii}(r,r')=N\int\Psi^{i}(r,r_2,r_3,...,r_N)\Psi^{i}(r',r_2,r_3,...,r_N)dr_2...dr_N$, where $N$ is the number of electrons and $\Psi^{i}$ is the generalized Kohn-Sham excited-state wavefunction that consists of a sum of Slater determinants of generalized Kohn-Sham orbitals with coefficients obtained from TDDFT \cite{Plasser2014}. To quantify the magnitude of charge transfer we integrated over subsystem difference densities. For this purpose, we subdivided the volume containing the difference densities of our full model systems into subsystem volumes, each containing one pigment. Note that $P$ is separated into $P_A$ and $P_B$, to enable the characterization of internal charge-transfer states of the type $P_A^+P_B^-$. Our aim is to assign each grid point of the difference-density grid to its closest pigment molecule. For achieving this, we tested two methods for assigning grid points to subsystem volumes: In method 1, we used the distances between grid points and each molecule's atomic coordinates (including hydrogen atoms). In method 2, we used distances between grid points and each molecule's geometrical center of gravity. Both methods result in the same trends, although the absolute values of the integrated subsystem densities differ slightly. 

\section{Results and Discussion}
\subsection{Absorption Spectrum and Excited State Character of the Bare Hexameric RC Model} \label{sec:spectrumRC}
\begin{figure}[htb]
		\centering
			\includegraphics[width=0.40\textwidth]{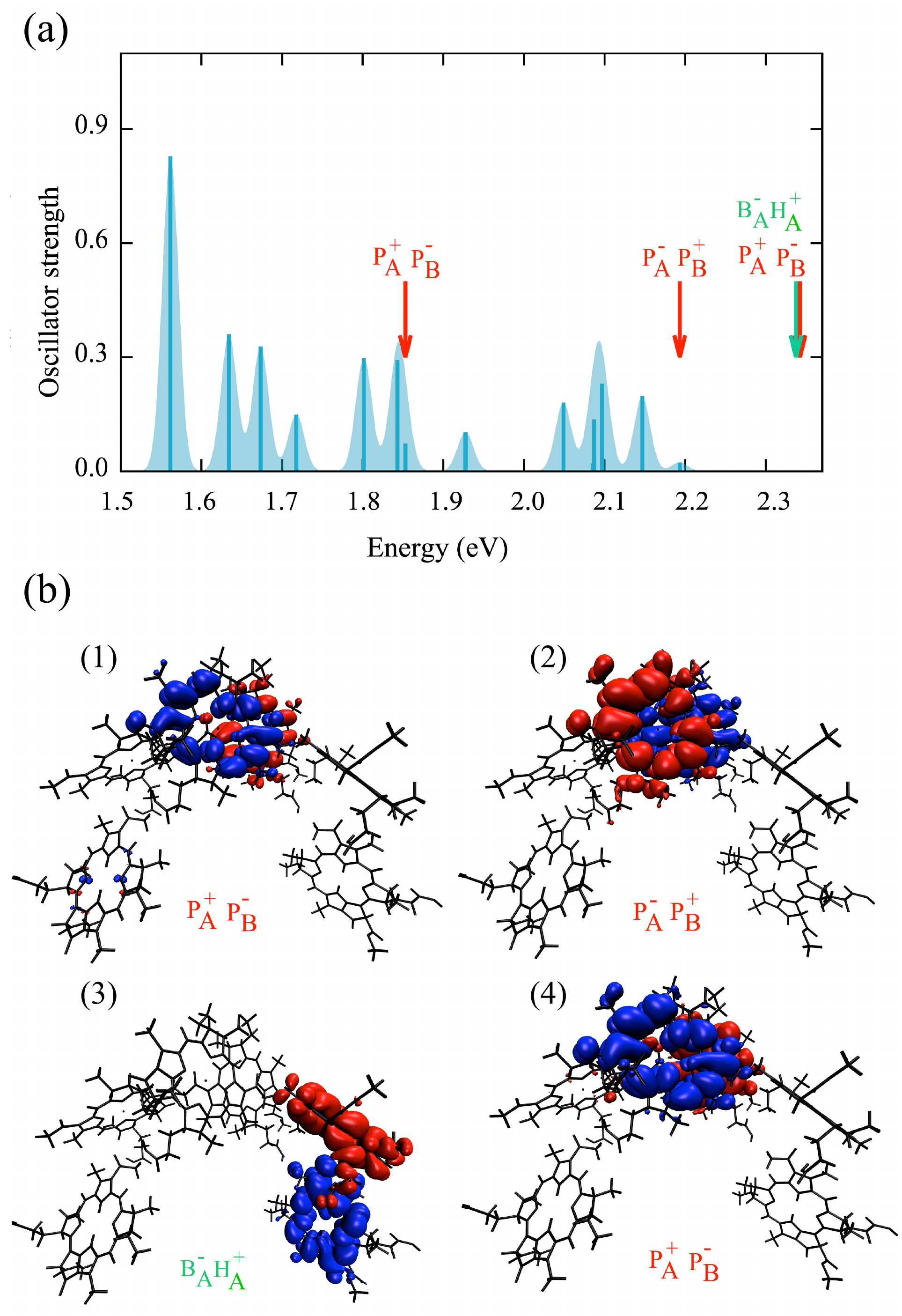}
		\caption{(a) TDDFT absorption spectrum of the bare hexameric RC model. Arrows mark excitations with low/vanishing oscillator strength and (partial) charge-transfer character. The shaded areas are calculated by folding the excitation energies with Gaussian functions with a width of 0.08\,eV as a guide to the eye. (b) Difference densities of the four charge-transfer excitations in this energy range.}
		\label{fig:spectrumRC}
  \end{figure}

We start our discussion by inspecting the absorption spectrum of a hexameric model of the RC based on the crystal structure as described in Section~\ref{sec:structures} and without including any parts of the environment, shown in Figure~\ref{fig:spectrumRC}a. For this model, we were able to calculate 16 excitations, corresponding to the energy range depicted in Figure~\ref{fig:spectrumRC}a. This energy range is dominated by $Q$-band excitations, i.e., excited states that originate from the coupling of the $Q_y$ and $Q_x$ excitations of the individual BCL and bacteriopheophytin molecules. However, because of the spatial proximity of these pigments in the RC, not all excitations can clearly be classified as coupled $Q_y$ or $Q_x$ as apparent from their transition densities shown in Figure S2. These transition densities also show that the majority of $Q$-band excitations is spatially strongly delocalized across several pigments, with some of them spreading over the entire RC model. This is the first main result of our study. A list of excitation energies, oscillator strengths, and spatial character as determined from the transition densities (and difference densities in case of charge-transfer excitations) can be found in Table~\ref{tab:excitations}. In this Table and in the rest of the text, the notation $(PBH)^*$ corresponds to an excitation delocalized across $P$, $B_A$, $B_B$, $H_A$, and $H_B$, while $P_A^+P_B^-$ denotes a charge-transfer excitation from $P_A$ to $P_B$.
    \begin{table}[htb]
    	\begin{tabular}{c|c c c}
    		\hline 
    		\hline 
    		\# & Energy & Oscillator Strength & Character  \\ 
    		\hline 
  1& 1.56  & 0.83  &  (PBH)$^*$ \\
  2& 1.63  & 0.36  &  (PBH$_{B}$)$^*$\\
  3& 1.67  & 0.33  &  (PBH)$^*$\\
  4& 1.71  & 0.15  &  (PBH)$^*$\\
  5& 1.80  & 0.30  &  (BH$_{B}$)$^*$ \\
  6& 1.84  & 0.29  &  (PB$_{A}$H)$^*$ \\
  7& 1.85  & 0.07  &  P$_{A}^{+}$P$_{B}^{-}$ \\
  8& 1.92  & 0.10  &  (PBH)$^*$ \\
  9& 2.04  & 0.18  &  (PBH)$^*$\\
 10& 2.08  & 0.02  &  (P$_{B}$BH$_{B}$)$^*$ \\
 11& 2.08  & 0.13  &  (PBH)$^*$ \\
 12& 2.09  & 0.23  &  (PBH$_{A}$)$^*$\\
 13& 2.14  & 0.20  &  (PBH$_{B}$)$^*$  \\
 14& 2.19  & 0.02  &  P$_{A}^{-}$P$_{B}^{+}$  \\
 15& 2.33  & 0.00  &  B$_{A}^{-}$H$_{A}^{+}$ \\
 16& 2.34  & 0.00  &  P$_{A}^{+}$P$_{B}^{-}$\\
		\hline  		
    	\end{tabular}
    	\caption{Excitation energies (in eV), oscillator strengths, and spatial delocalization/charge-transfer character of the first 16 excitations of our bare hexameric RC model structure}
     \label{tab:excitations}
    \end{table} 

We find four excitations with charge-transfer character in this energy range. The difference densities of these excitations are depicted in Figure~\ref{fig:spectrumRC}b (all other differences densities for this structural model can be found in Figure S3). Here and in the following, positive difference density values indicate a region of space in which the electron density (i.e., negative charge density) increases as a consequence of the excitation (shown in blue), whereas negative values indicate regions of space in which the electron density decreases (shown in red). Numerical values based on the integration of difference densities as described in Section~\ref{sec:CT} are listed in Table S3. The three charge-transfer excitations within the special pair, corresponding to $P_A^+P_B^-$ and $P_A^-P_B^+$ arise as a consequence of the strong coupling of $P_A$ and $P_B$. In particular the first $P_A^+P_B^-$ excitation mixes strongly with other excitations at $\sim$1.85\,eV and therefore exhibits partial charge-transfer character, in which 0.69 of an electron is transferred from $P_A$ to $P_B$. Strikingly, we also find a charge-transfer excitation from $B_A$ to $H_A$ in this energy range. This is the lowest-energy pure charge-transfer state we find in our calculations (0.99 of an electron is transferred from $B_A$ to $H_A$). The second main result of our study is that the appearance of this charge-transfer state at $\sim$2.3\,eV is a consequence of the spatial arrangement of the pigments in the bacterial RC alone. We will discuss how the energy of this state is affected by including environmental effects in Section~\ref{sec:Abranch}.

\subsection{Effect of the Protein Environment on a Tetrameric RC Model}
\label{sec:tetrameric}
The importance of the protein environment and its impact on charge transfer were recognized already in early studies of the bacterial RC \cite{Stanley1996,Steffen1994,Hiyama2011}. Proposals for how the surrounding proteins affect charge transfer in the RC have primarily included asymmetries in the dielectric environment and in protein electrostatic fields that $A$ and $B$ branch cofactors experience \cite{Lockhart1990,Steffen1994,Alden1995,Gunner1996,Saggu2019}. Our goal here is to include parts of the protein environment explicitly in our TDDFT calculations to elucidate which amino-acid residues electronically couple to the primary RC pigments. For this purpose, we start by studying tetrameric models of the RC including the special pair $P$ and the accessory BCLs $B_A$ and $B_B$, and systematically increase the number of amino-acid residues in our calculations.
\begin{figure*}[htb]
		\centering
		\includegraphics[width=0.6\textwidth]{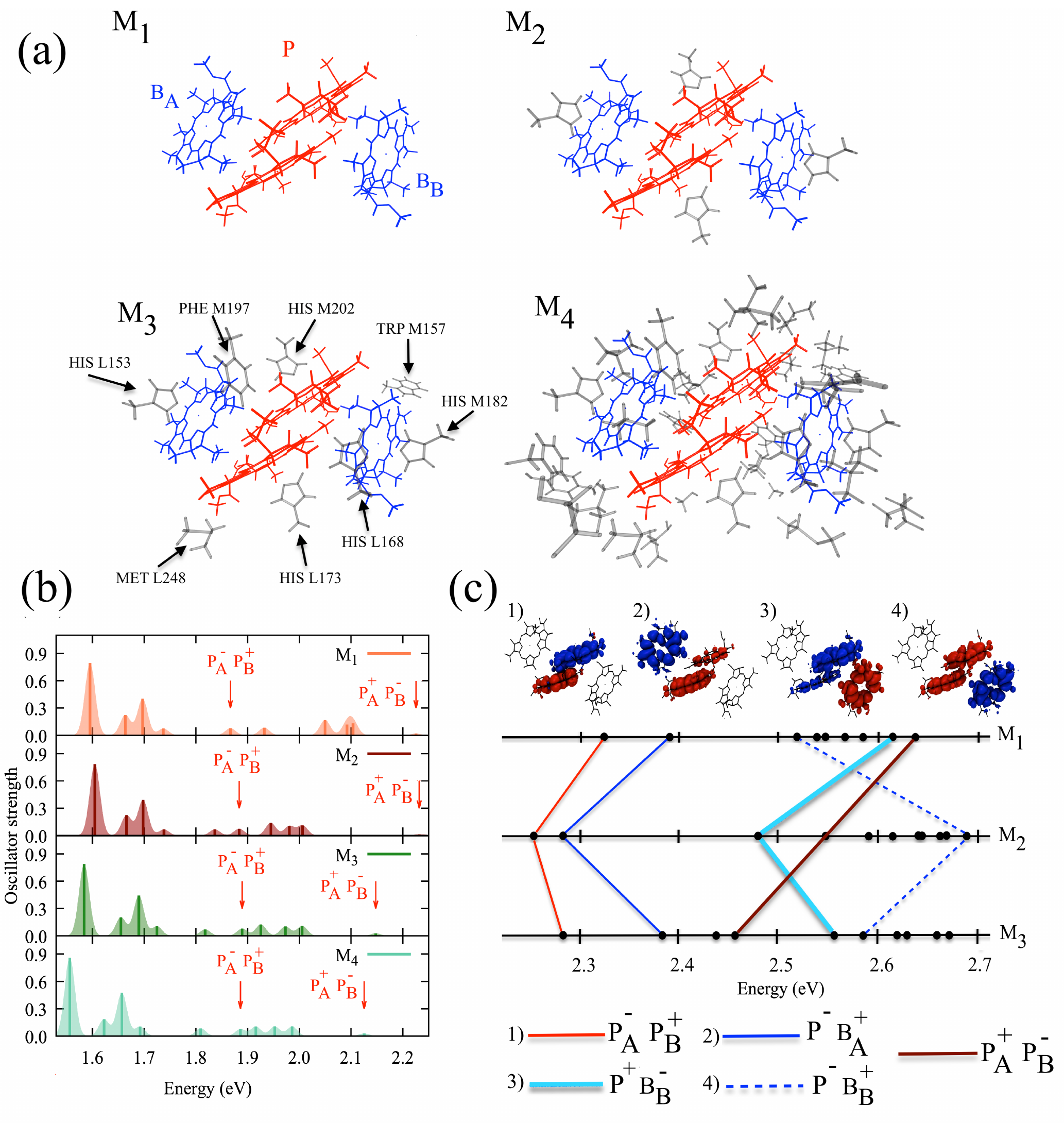}
		\caption{(a) Representation of model systems $M_1$, $M_2$, $M_3$, and $M_4$ as described in the main text. (b) Absorption spectra of $M_1-M_4$ in the energy region where coupled $Q_y$ and $Q_x$ excitations are expected. Red arrows mark excitations with charge-transfer character. (c) Energy of dark excitations (zero oscillator strength) of $M_1 - M_3$ and difference densities of selected charge-transfer excitations of $M_1$. The red surface of the difference density shows the isovalue -0.0001$a_0^{-3}$ and the blue surface shows 0.0001$a_0^{-3}$.}
\label{fig:spectra}
\end{figure*}

We construct four model systems shown in Figure~\ref{fig:spectra}a: Model system $M_1$ consists of the four BCL molecules $P_A$, $P_B$, $B_A$, and $B_B$. For a direct analysis of the influence of the closest lying amino acids, the histidine molecules that coordinate each of the BCLs (HIS M202, HIS M182, HIS L173, HIS L153) were included in model system $M_2$. Our largest model system, $M_4$, contains \textit{all} amino acids in a radius of $3\text{\AA}$ around the BCLs. These 32 amino acids were determined by constructing spheres with radius $3\text{\AA}$ around each atom of the four BCL molecules (excluding hydrogen atoms and the phytyl tail). A complete list (Table S4) and all structure files can be found in the SM. We calculated the electronic density of states (DOS) of these model systems and found two occupied states localized on amino acids TRP M157 and MET L248, respectively, energetically close to the highest occupied molecular orbital of $M_4$ (see Figures S4 and S5). However, a model system $M_3^*$ consisting of the four primary BCL molecules, the coordinating histidines, and these two amino acids features an electronic DOS distinctly different from that of $M_4$ (Figure S6). We therefore additionally included the two main symmetry breaking amino acids PHE M197 and HIS L168 as suggested by Eisenmayer \textit{et al.}~\cite{Eisenmayer2013} to construct model system $M_3$ with a DOS in very good agreement to the DOS of $M_4$ in the relevant energy range.

The TDDFT $Q$-band spectra of $M_1$ -- $M_4$ comprising the ten lowest-energy excitations are shown in Figure~\ref{fig:spectra}b and Tables S5 and S6. The first four excitations of these model systems can be seen as arising from a coupling of the $Q_y$ excitations of $P$, $B_A$ and $B_B$. We provide a detailed analysis of the origin of these excitations in the SM (Figures S8 -- S10). Inspection of their transition densities (Figure S8) shows that only the first excitation is localized on $P$ while excitations 2 to 4 are coupled $Q_y$ excitations spread across all four BCLs. Among the following six excitations of $M_1$, 5, 6, and 7 can clearly be assigned to $P$. Two of these excitations (6 and 7) have coupled $Q_x$ character; excitation 5 has $Q_y$ character but integration over the difference density corresponding to this state also shows substantial charge-transfer character. States 8 and 9 of $M_1$ are $Q_x$ excitations associated with $B_A$ and $B_B$. Excitation 10 of $M_1$ is nearly dark and corresponds to a $P_A^+P_B^-$ charge-transfer state.

Inclusion of the histidines in $M_2$ hardly affects the first four excitations. Only when further amino-acid residues are added do we observe a noticeable redshift: the first excitation of $M_4$ is 40\,meV lower in energy than that of $M_1$. The character of these excitations is not changed by the environment. The average difference density of excitations 1-4 is barely affected by addition of the environment as shown in Figure S11. We observe more significant energy differences for the next six excitations. Excitations with $Q_x$ character are redshifted by $\sim$100\,meV through addition of the coordinating histidines and the coupled $Q_x$ excitations of $P$ are redshifted by another 20\,meV for model system $M_3$, while the $Q_x$ excitations of the accessory BCLs are stable. We note that the significance of the coordinating histidines for these excitations can also be seen in the difference density (Figure S11). In systems $M_2$ - $M_4$, there is a clear transfer of positive charge from $P$ to the coordinating histidines.

The energy of the (partial) charge-transfer excitations 5 and 10 of $M_1$ are also affected by adding the protein environment. Excitation 5 of $M_1$ is redshifted by $\sim$60\,meV and becomes excitation 6 in $M_4$, and excitation 10 is redshifted by $\sim$100\,meV. The magnitude of charge transfer is only slightly affected by the addition of the environment. The amount of charge transferred from $P_A$ to $P_B$ in excitation 5 of $M_1$ decreases slightly when adding the environment while in excitation 10 it remains the same. Overall, the character of excitations 5 to 10 is only slightly affected by the amino-acid environment, as can be seen in Figure S12, which shows the average difference density of excitations 5-10 for systems $M_1$ - $M_4$. Importantly, Figure~\ref{fig:spectra}b demonstrates that the spectrum of $M_4$ is very similar to that of $M_3$ with the same order of states and only a small global redshift as compared to $M_3$. We therefore conclude that the eight amino acids considered in $M_3$ reproduce the main (static) effects of the amino-acid environment. Therefore, they constitute a "reasonable minimal environment" that should be included explicitly in future calculations.

At energies above the $Q$-band and below the Soret band that starts at $\sim$3\,eV\cite{Cogdell2006}, we observe a range of dark states with charge-transfer character. Here, we only discuss five states in this energy region that lead to a clear charge transfer between different BCL molecules. The energy and difference densities of these states are shown in Figure~\ref{fig:spectra}c. (Integrated) difference densities of all states are shown in Figure S13 and Table S7. Due to the large size of $M_4$, we only calculated twelve excitations with high numerical accuracy for this system. Since the effects of the environment are well-represented by $M_3$ as shown before, we do not discuss the charge-transfer excitations of $M_4$ in detail. However, Table S5 and S6 show that the energies of the charge-transfer excitations 11 and 12 of $M_3$ and $M_4$ are in very good agreement.

In similarity to the findings for our hexameric model of the RC, we find a $P_A^-P_B^+$ excitation in $M_1$-$M_3$ (excitation 11). Addition of the histidines redshifts this state by 80\,meV, while the additional amino acids of $M_3$ lead to a blueshift of 26\,meV. The 12th excitation corresponds to $P^+B_A^-$. This state is redshifted by 120\,meV through the addition of histidines while further amino acids blueshift the excitation back by 110\,meV. The energy gap between $P^+B_A^-$ and the next charge-transfer excitation is substantial: $\sim$130\,meV in $M_1$, $\sim$210\,meV in $M_2$ and 85\,meV in $M_3$. In the 13th excitation of $M_1$, we observe a backward charge transfer from the B branch corresponding to $P^-B_B^+$. The first forward charge transfer into the B branch ($P^+B_B^-$) occurs at $\sim$2.6\,eV, i.e., at significantly higher energies than $P^+B_A^-$. The addition of amino acids affects this excitation in a similar way as $P^+B_A^-$. These calculations show that inclusion of a (static) protein environment for the system studied here changes excitation energies substantially but hardly affects the character and spatial delocalization of states. For our tetrameric model systems addition of the protein environment also does not lead to a mixing of $Q$-band excitations and experimentally relevant charge-transfer states.

To approximate the effect of molecular vibrations on the energy of the excited states, we also calculated the normal mode spectrum of model system $M_1$ in \textsc{turbomole} using the B3LYP functional and def2-SVP basis set. We then distorted the structure along each of the normal modes with a distortion amplitude corresponding to 300\,K, and calculated the TDDFT excitation spectra in \textsc{qchem} with $\omega$PBE as before. The high-frequency modes of $M_1$ correspond to intramolecular vibrations such as C-C and C-H stretch modes, which are not thermally activated and only have a small effect on the energy of the delocalized excitations that we are interested in here. We therefore only calculated the effect of normal mode distortions on excitation energies for wavenumbers below 85\,cm$^{-1}$. Low-frequency modes correspond to intermolecular vibrations that change the orbital overlap between neighboring BCL molecules and are thus expected to have a more substantial effect on excitation energies of delocalized and charge-transfer excitations \cite{alvertis_impact_2020}.

Our results, shown in Figure S14, confirm this intuitive picture: The first excitation, corresponding to a coupled $Q_y$ excitation of $P$, exhibits mode-dependent energy changes of up to 20\,meV, while excitations 2 -- 4 are much less sensitive to these distortions with energy changes of $\lesssim$10\,meV in line with their spatial delocalization across $B_A$ and $B_B$, which are far apart. A similar observation holds for the coupled $Q_x$ excitations of $P$, $B_A$ and $B_B$. On the other hand, excitations with (partial) charge-transfer character between neighboring molecules are highly sensitive to low-frequency vibrations, exhibiting excitation-energy changes of up to 30\,meV for charge transfer between $P_A$ and $P_B$, and up to 50\,meV for charge transfer between $P$ and $B_A$ or $B_B$. These excitation-energy changes can result in both red- and blueshifts, as shown in Figure S14. Nonetheless, our analysis indicates that inclusion of inter- and intramolecular vibrations does not change our overall result that in a tetrameric model of the bacterial RC, charge-transfer excitations into the A-branch are energetically well-separated from the $Q$-band excitations. This finding is in line with systematic first-principles molecular dynamics simulations of the RC of the heliobacterium \textit{Heliobacterium modesticaldum} in Ref.~\citenum{Brutting2021}. While the first A-branch charge-transfer state is at considerably lower energies for that system, thermal vibrations do not significantly change the spectral overlap of the charge-transfer state with the coupled $Q_y$ excitations as compared to the static structural model \cite{Brutting2021}.

Finally, we tested whether the inclusion of further parts of the protein environment through a QM/MM scheme would change our main conclusions. Figure S15 and S16 show the full TDDFT spectrum and the TDA spectrum of $M_1$ with and without the QM/MM environment. Inclusion of the QM/MM environment leads to changes in the absorption spectrum of comparable size as in our explicit model $M_4$. In particular, we also observe a redshift of the first charge-transfer excitation of $\sim$200\,meV. However, the redshift of the coupled Q$_y$ and Q$_x$ excitations due to the protein environment is smaller in the QM/MM model, and some of the detailed changes in the partial and full charge-transfer excitations are also not captured by the MM environment.

\subsection{Effect of Environment on A and B Branch Excited States}\label{sec:Abranch}
Due to the large size of the hexameric model of the RC discussed in Section~\ref{sec:spectrumRC} (494 atoms), a TDDFT calculation including the relevant charge-transfer states for a structural model that also includes significant parts of the protein environment as done for the tetrameric model in Section~\ref{sec:tetrameric} is computationally not feasible. The largest hexameric RC model that we could run full TDDFT calculations for includes the coordinating histidines close to $P_A$, $P_B$, $B_A$ and $B_B$ (as in $M_2$) plus two leucines close to $H_A$ and $H_B$. Table S2 shows that including these amino-acid residues has similar effects as observed in Section~\ref{sec:tetrameric}, but the calculation of more than the first 13 excitations was not feasible for this system.

\begin{figure}[htb] 
	\centering
		\includegraphics[width=0.45\textwidth]{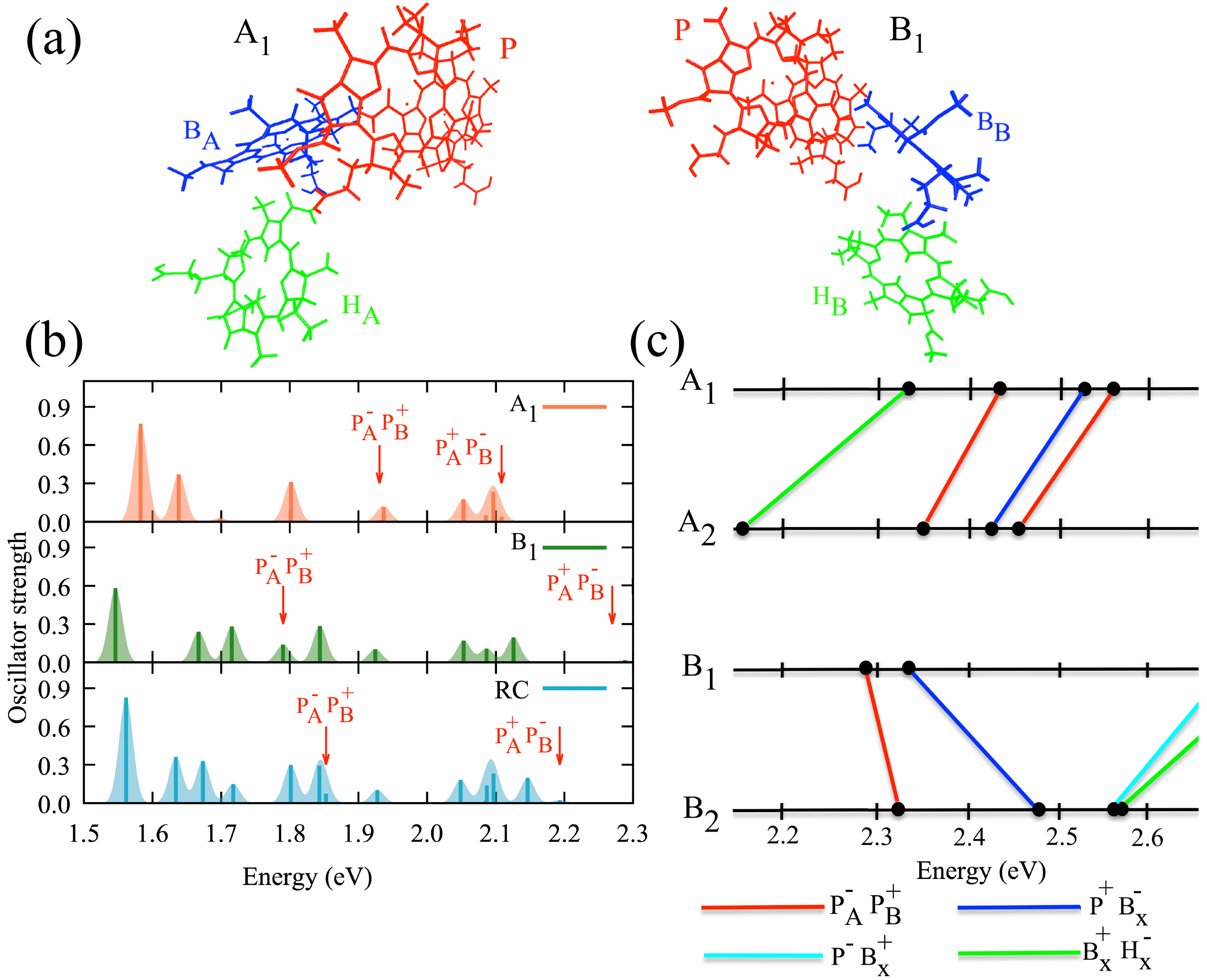}
\caption{(a) Structures of model systems $A_1$ (including $SP$+$B_A$+$H_A$) and $B_1$ including $SP$+$B_B$+$H_B$ . (b) Absorption spectra of $H_{A1}$, $H_{B1}$ and $RC$ in the energy region where coupled $Q_y$ and $Q_x$ excitations are expected. Red arrows mark excitations with charge-transfer character. (c) Energy of dark excitations (zero oscillator strength) of $A_1 - A_2$ and $B_1 - B_2$.}
\label{fig:ABbranch}
\end{figure}

However, given the large spatial separation between A- and B-branch accessory BCLs and bacteriopheophytins, we can assume that parts of the spectrum of the full RC arise as combinations of A-branch and B-branch excitations, respectively. To test this assumption, we constructed two further structural models $A_1$ and $B_1$, shown in Figure~\ref{fig:ABbranch}a, comprising $P$, $B_A$, and $H_A$ for the A branch, and $P$, $B_B$ and $H_B$ for the B branch, respectively. We compare the excitation spectrum of the hexameric model with that of $A_1$ and $B_1$, respectively, in Figure~\ref{fig:ABbranch}b. As expected, excitations associated with $P$ appear in all three spectra, albeit at different energies (e.g., the first excitation and the charge-transfer states $P_A^-P_B^+$ and $P_A^+P_B^-$). On the other hand, excitations that are localized on the A- or B-branch, can clearly be assigned to either $A_1$ or $B_1$. In particular, in our calculation of the spectrum of $A_1$, we find the charge-transfer state $B_A^+H_A^-$ at the same energy ($\sim$2.3\,eV) as in our hexameric model. We can therefore study the effect of adding amino acids to the energy of this and other relevant charge transfer states using our structural models $A_1$ and $B_1$ as a starting point. 

A comparison of the excitation energies of the relevant charge-transfer states in A- and B-branch is shown in Figure~\ref{fig:ABbranch}c (and listed in Tables S9 and S10). The $B_A^+H_A^-$ state is significantly lower in energy than $B_B^+H_B^-$, in agreement with the experimentally observed directionality of charge-separation along the A-branch. By adding the coordinating histidines and leucines in model systems $A_2$ and $B_2$, both states are redshifted by more than 100\,meV. Charge-transfer states from $P$ to $B_B$ and $P$ to $B_A$ are significantly higher in energy, in particular after addition of the coordinating histidines and leucines. Adding further amino-acid residues (listed in Table S11), in analogy with $M_3$ in Section~\ref{sec:tetrameric}, leads to a further redshift of $B_A^-H_A^+$, bringing this charge-transfer state within 25\,meV of the coupled $Q_x$ excitations (see Table S9). It is therefore likely that addition of further parts of the protein environment in concert with thermally-activated molecular vibrations could lead to a mixing of this charge-transfer state and the delocalized coupled $Q_x$ excitations. 
    
\section{Summary and Conclusions}
\label{summary_section}

Our first principles calculations show a pronounced effect of the protein environment on the electronic structure and excited states of the primary six pigments comprising the RC of \textit{Rhodobacter sphaeroides}. By systematically adding relevant amino acids in the vicinity of these chromophores, we find a significant redshift of the coupled $Q_y$ and Q$_x$ excitations. Charge-transfer excitations are observed in the form of dark excitations starting at $\sim$0.2\,eV above the coupled $Q_x$ excitations. These charge-transfer states are strongly affected by direct inclusion of the protein environment with energy changes of up to $\sim$0.2\,eV. However, contrary to the coupled $Q_y$ and $Q_x$ excitations, the protein environment affects charge-transfer states of different character differently. In particular, the lowest-energy charge transfer state in our calculations corresponds to $B_A^+H_A^-$ and is significantly lower in energy than other excitations that move charge into the A branch. It is also almost 0.5\,eV lower than an equivalent excitation in the B-branch. The $B_A^-H_A^+$ excitation is redshifted by the inclusion of close-lying amino-acid residues and can mix with the coupled $Q_x$ excitations.

Beyond suggesting a possible mechanism for charge-transfer in the RC of \textit{Rhodobacter sphaeroides}, our calculations demonstrate the complex excited state landscape of the RC's chromophores. Analysing the transition and differences densities of the excited states allows for several conclusions. While most of the $Q$-band excitations can be understood as a consequence of the coupling of $Q_y$ and $Q_x$ excitations of the special pair BCLs $P$ and the accessory BCLs $B_A$ and $B_B$, the close spatial proximity of these molecules leads to strong coupling, mixing in (partial) charge-transfer states of the type $P_A^+P_B^-$ at relatively low energies. Furthermore, only the first high-oscillator strength excitation of the spectrum can be interpreted as a coupled $Q_y$ excitation of the special pair. All other excitations are strongly delocalized, some of them with significant transition density on all six primary pigments. The presence of strongly delocalized excited states corresponding to both energy- and charge-transfer excitations is relevant because delocalized excitations are expected to be more strongly affected by thermally activated molecular vibrations than localized excitations as shown previously by Alvertis \textit{et al.} for the oligo-acene series \cite{alvertis_impact_2020}. In our study we have approximately shown this effect by calculating the exciton renormalization energies of tetrameric model structures distorted along vibrational normal modes. Based on our own calculations and previous literature \cite{Eisenmayer2012, Eisenmayer2013, Tamura2021, Mitsuhashi2021}, we expect yet more pronounced effects for larger structural models that also include the vibronic coupling to the protein environment. 

Finally, our calculations allow us to comment on the interpretation of experimental spectroscopy of pigment-protein complexes like the bacterial RC. Our results suggest that because of the delocalized nature of energy- and charge-transfer excitations in these systems, the assignment of spectroscopic features to linear combinations of localized excitations on individual pigments is not always justified. Care should be taken when modelling the strongly coupled pigments of the bacterial RC in terms of their constituting elements.

\section*{Supplementary Material}
Additional convergence data, excitation energies, difference densities and transition densities not shown in the main text, a discussion of the spectral origin of the coupled $Q_y$ and $Q_x$ excitations, details on the QM/MM calculations, results for vibrationally excited structures, and structure files of model systems $M_1$ - $M_4$ and our bare hexameric model.

\begin{acknowledgements}
This work was supported by the Bavarian State Ministry of Science and the Arts through the Elite Network Bavaria (ENB), the Collaborative Research Network Solar Technologies go Hybrid (SolTech), the Study Program "Biological Physics" of the ENB, and through computational resources provided by the Bavarian Polymer Institute (BPI).
\end{acknowledgements}

\end{document}